\documentstyle[aps]{revtex}
\begin{document}
\draft
\input{psfig}
\title{Tunnel junctions of unconventional superconductors}
\author{C. Bruder}
\address{Theoretische Festk\"orperphysik,
	Universit\"at Karlsruhe, D-76128 Karlsruhe, FRG}
\author{A. van Otterlo\footnote{E-mail address: avo@itp.phys.ethz.ch}}
\address{Theoretische Physik, ETH-H\"{o}nggerberg,
	CH-8093 Z\"{u}rich, Switzerland}
\author{and G.~T. Zimanyi}
\address{Department of Physics, University of California,
	Davis, CA 95616, USA}
\maketitle
\begin{abstract}
The phenomenology of Josephson tunnel junctions between unconventional
superconductors is developed further. In contrast to s-wave
superconductors, for d-wave superconductors the direction
dependence of the tunnel matrix elements that describe the barrier
is relevant. We find the full I-V characteristics and comment on the
thermodynamical properties of these junctions. They depend sensitively
on the relative orientation of the superconductors. The I-V
characteristics differ from the normal s-wave RSJ-like behavior.
\end{abstract}
\pacs{PACS numbers: 74.50.+r, 74.20.-z}

The symmetry of the superconducting order parameter of the high-T$_C$
compounds is the subject of a heated debate. Quite early there
were theoretical suggestions \cite{maurice,scalapino,kotliar,claudius}
that these materials are unconventional superconductors that have
$d_{x^{2}-y^{2}}$-symmetry of the order parameter. They have stimulated
both experimental and theoretical work.
The experimental situation is still unclear \cite{phystoday}. Several
experiments indicate d-wave pairing, e.g., the temperature dependence
of the penetration depth $\lambda(T)$ \cite {hardy} and the NMR relaxation
rates \cite{martindale,pines}. The most convincing set of experiments up
to now are the SQUID experiments \cite{wollman,brawner}, which show that
a SQUID loop consisting of a high-T$_C$ crystal closed by an ordinary
s-wave superconductor shows a phase shift of the order of $\pi$ in the
dependence of the critical current on the flux, and the experiments on
flux quantization in multi-domain rings \cite{tsuei,kirtley}.
There are other experiments,
however, that cannot be reconciled with the $d_{x^{2}-y^{2}}$-model and
have been interpreted to indicate (possibly anisotropic) s-wave pairing
\cite{chaudhari,sun}, see, however, Ref. \cite{millis}.
Since a microscopic theory is still lacking, there are also different
theoretical opinions on the symmetry of the order parameter. Some
phenomenological calculations, assuming d-wave pairing, have been done
and they yield partial understanding of some experimental data.

\begin{figure}
\centerline{\psfig{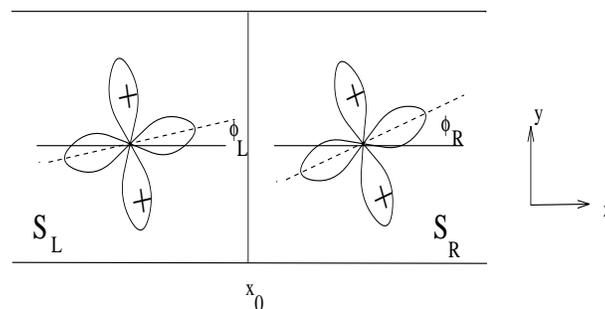}}
\caption{The system we consider consists of two coupled d-wave
superconductors. Their orientation is characterized by the angles
$\phi_{L}$ and $\phi_{R}$.
}
\label{fig1}
\end{figure}

In this note we will further develop the phenomenology of unconventional
superconductors, assuming BCS-like behavior for the density of states.
Our results are not specific for high T$_{C}$ materials, but may be
relevant also for other unconventional superconductors such as heavy fermion
compounds.
We will consider tunnel junctions between two pieces of bulk superconductor
as shown in Fig. \ref{fig1}. The superconductor is assumed to have a
$k$-dependent order parameter of the form
$\Delta_{k}=\Delta_{0}[\hat{k}^{2}_{x}-\hat{k}^{2}_{y}]$.
In contrast to earlier work \cite{yip,maki,tanaka,xu}, we include
the direction dependence of the tunnel matrix elements
that describe tunneling across the insulating barrier between the
superconductors. For s-wave superconductors this dependence drops out, but
for d-wave superconductors it is essential \cite{rainer,zaikin}.

Along the lines of Ref. \cite{ab} we calculate the full current-voltage
characteristic
(I-V curves) for different relative orientations. We find that the presence
of gap-nodes strongly influences the I-V characteristics. For a specific
relative orientation the quasi-particle current will be proportional
to the square of the voltage, which leads to new behavior for the total
current, which is different from the usual RSJ-like overdamped junction.
We will also comment on the thermodynamic properties \cite{aes} of
d-wave tunnel junctions  and the relevance of our work for thin
granular high-T$_{C}$ films \cite{films}.

A convenient starting point is the tunneling Hamiltonian
\begin{equation}
	H=H_{L}+H_{R}+\int_{r\in L}\int_{r'\in R}\sum_{\sigma}
	[T(r,r')\psi^{\dagger}_{r,\sigma}\psi_{r',\sigma}+h.c.]\;.
\end{equation}
Here $H_{L}$ and $H_{R}$ denote the unperturbed Hamiltonians of
the left and right superconductor. From a second order perturbation
expansion in the tunnel matrix elements one finds the
Ambegaokar-Baratoff \cite{ab} formula that expresses the current
across a tunnel junction in second order perturbation theory
in the tunnel matrix elements as
\begin{eqnarray}
	I=2e\;\mbox{Im} \int\frac{d^{3}k}{(2\pi)^3}\int
	\frac{d^{3}k'}{(2\pi)^3}\int\frac{d\omega}{2\pi}
	\int\frac{d\omega'}{2\pi}\left[f_{L}(\omega)-
	f_{R}(\omega')\right]
	\nonumber\\
	\left\{\mid T_{k,k'}\mid^{2}
	\frac{A(k,\omega)A(k',\omega')}{\omega-\omega'+i\eta}+
	T_{k,k'}T_{-k,-k'}\frac{\bar{B}(k,\omega)B(k',\omega')}
	{\omega-\omega'+i\eta}e^{i(\varphi+2(\mu_{L}-\mu_{R})t)}
	\right\}\;.
	\label{curr}
\end{eqnarray}
Here $A$ and $B$ are the spectral densities of the normal and anomalous
Greens functions, $\varphi$ is the phase difference between the two
superconductors, and $f_{L/R}$ the Fermi distributions at the chemical
potential $\mu_{L/R}$. The left and right chemical potential differ by
the applied voltage, $\mu_{L}-\mu_{R}=eV$. The $T_{k,k'}$ are the matrix
elements that transfer electrons from a state $k$ in one superconductor
to a state $k'$ in the other. The first term in (\ref{curr}) describes
the quasi-particle current
$I_{QP}$ and the second term the supercurrent which is proportional to
the critical current $I_{CR}$. The spectral
densities have the form
\begin{eqnarray}
	A(k,\omega)&=&\pi\left[\left(1+\frac{\epsilon_{k}}{E(k)}\right)
	\delta(\omega-E(k))+\left(1-\frac{\epsilon_{k}}{E(k)}\right)
	\delta(\omega+E(k))\right]
	\nonumber\\
	B(k,\omega)&=&\pi\frac{\Delta(\hat{k})}{E(k)}
	\left[\delta(\omega+E(k))-\delta(\omega-E(k))\right]\;,
	\nonumber
\end{eqnarray}
where $E(k)=\sqrt{\mid \Delta(\hat{k}) \mid^{2}+\epsilon^{2}_{k}}$.
The strategy is to take seriously the orientation dependence
\cite{rainer,zaikin}.
This is necessary, since with the standard assumption that the tunnel
matrix elements $T_{k,k'}$ are independent of momenta, Eq.
(\ref{curr}) yields a critical current $I_{CR}=0$ after an angular
average with $\Delta_{k}=\Delta_{0}[\hat{k}^{2}_{x}-\hat{k}^{2}_{y}]$.

Thus, in order to obtain physical results, the direction
dependence of tunneling has to be taken into account. We will
consider the case of a smooth barrier for which momentum parallel
to the barrier is conserved during tunneling.
A reasonable assumption seems to be to take the tunnel matrix
elements $T(r,r')$ non-vanishing only when $r$ and
$r'$ are both close to the barrier. We take
\begin{equation}
	T(r,r')=\tilde{T}\delta^{(3)}(r-r')\delta^{(1)}(r_{x}-x_{0}),
\end{equation}
where $x_{0}$ is the location of the barrier, as indicated in
Fig. \ref{fig1}. This ensures that the momentum parallel
to the tunnel barrier is conserved, since the Fourier transform
of the matrix element is
\begin{equation}
	T_{k,-k'}=(2\pi)^{2}\tilde{T}\delta^{(2)}(k_{\|}-k'_{\|})
	f(\mid\hat{k}_{x}\mid)\Theta(k_{x}\cdot k'_{x})
	\label{tunn}
\end{equation}
The last factor is a restriction on the direction of tunneling. Before
tunneling the electron should move towards the barrier, after tunneling
away from the barrier. The function $f(\mid\hat{k}_{x}\mid)$ is
a weight function
that makes tunneling perpendicular to the barrier more probable than tunneling
parallel to the barrier. If one models the barrier by a slab of finite
thickness and takes the tunneling probability to be exponentially small in
the traversed distance in the barrier, one deduces
$f(\mid\hat{k}_{x}\mid)\sim \exp(-1/\mid\hat{k}_{x}\mid)$. Numerical constants
may depend on the exact choice for $f$, but principal behavior should not
depend on it. In the following we take it to be
$f(\mid\hat{k}_{x}\mid)=\mid\hat{k}_{x}\mid$ which corresponds to vanishing
thickness of the barrier.

\begin{figure}
\centerline{\psfig{figure=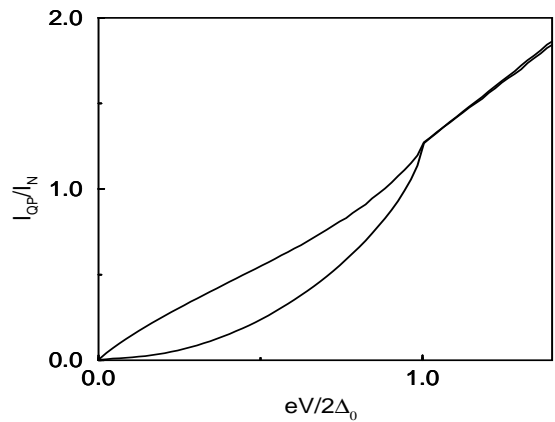,height=6cm,width=8cm}}
\caption{The quasi-particle current $I_{QP}$ as a function of
	the applied voltage for temperatures $T/\Delta_{0}$=1
	(upper curve), and 0.1 (lower curve) for the case
	$\phi_{L}=\phi_{R}$=0. Here and in the following figures,
	$I_N\equiv I_N(eV=2\Delta_0)$.}
\label{fig2}
\end{figure}

\begin{figure}
\centerline{\psfig{figure=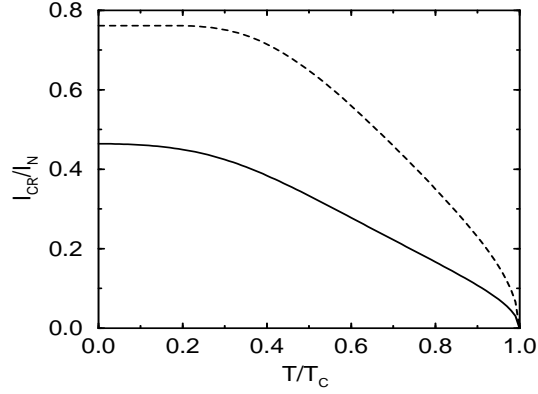,height=6cm,width=8cm}}
\centerline{\psfig{figure=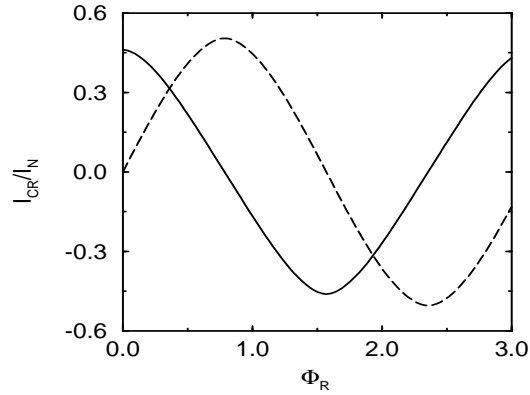,height=6cm,width=8cm}}
\caption{a) The critical current $I_{CR}$ as a function of
	temperature for the case $\phi_{L}=\phi_{R}$=0
	(full line). For comparison the s-wave result is also
	shown (dashed line).
	b) The critical current $I_{CR}$ as a function of the
	relative orientation $\phi_{R}$ ($\phi_{L}$=0 (full line), and
	$\phi_{L}=\frac{\pi}{4}$ (dashed line)) at $T$=0.}
\label{fig3}
\end{figure}

With this explicit form of the tunnel matrix elements, both the
quasi-particle current $I_{QP}$ as well as the critical current
$I_{CR}$ following from Eq. (\ref{curr}) can be evaluated by numerical
integration for different temperatures. For normalization we calculate
the normal state current for $\Delta_{0}$=0 with the help of (\ref{curr}).
This allows us to identify the normal state conductance $G_{N}$ of the
barrier as $G_{N}=8\pi^{5}\tilde{T}^{2}N(0)^{2}k^{-2}_{F}A/R_{K}$, where
$A$ denotes the area of the junction and $R_{K}=h/e^{2}$ the Klitzing
quantum of resistance. The expression for the critical current that we
find from Eq. (\ref{curr}) with $\mu_{l}=\mu_{r}$ is
\begin{equation}
	I_{CR}=2e \int d\omega d\omega'\int\frac{d\Omega}{4\pi}
	\frac{d\Omega}{4\pi}
	\frac{f(\omega)-f(\omega')}{\omega-\omega'}
	\frac{\Delta_{L}(\Omega)}{\omega}\frac{\Delta_{R}(\Omega')}{\omega'}
	N_{L}(\omega)N_{R}(\omega') \mid T(\Omega,\Omega')\mid^{2}\; ,
\end{equation}
where $N_{L}(\omega)=N(0)\mid\omega\mid\Theta(\mid\omega\mid-\Delta_L(\Omega))/
\sqrt{\omega^{2}-\Delta_L^{2}(\Omega)}$ denotes the (angle-resolved)
density of states of the left side, $N_R(\omega')$ is defined similarly
for the right side, $d\Omega=d\phi d\theta \sin\theta$, and
$\Delta_{L/R}(\Omega)=\Delta_{0}\cos[2(\phi-\phi_{L/R})]$.
The angles $\phi_{L}$ and $\phi_{R}$ determine the relative
orientation of the two superconductors. A natural choice
which applies to many junctions is $\phi_{L}=\phi_{R}=0$. The
results are shown in Figs. \ref{fig2} and \ref{fig3}a) for $I_{QP}$
and $I_{CR}$ respectively. The magnitude and sign of the critical
current show a strong orientation dependence, see Fig. \ref{fig3}b).

The quasi-particle current behaves mostly like for an s-wave
superconductor $I_{QP}\sim\exp\{-\Delta_{eff}/k_{B}T\}$, where
the leading behavior is determined by tunneling from a gap node
in one superconductor into the effective gap $\Delta_{eff}$ in
the other. However, for those special
relative orientations for which the gap nodes in the left and right
superconductor are parallel, the behavior is different. The
quasi-particle current $I_{QP}\sim V^{2}$ for voltages
$2\Delta_{0}> eV\geq k_{B}T$ and $I_{QP}\sim V$ for voltages $eV\leq
k_{B}T$. This can be understood as follows. The dominant contribution
to $I_{QP}$ arises from 'node to node' tunneling. The available phase
space around the gap-nodes scales with the bigger of $eV$ and $T$. With
the usual factor $V$ from the difference in $f_{L}-f_{R}$ this yields
the quoted behavior.

\begin{figure}
\centerline{\psfig{figure=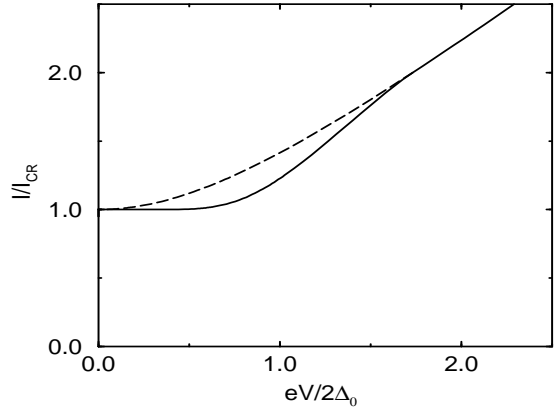,height=6cm,width=8cm}}
\caption{The complete I-V characteristic for a d-wave tunnel
	junction with $\phi_{L}=\phi_{R}=0$ at $T$=0 (full line).
	For comparison also the RSJ result is shown (dashed line).}
\label{fig4}
\end{figure}

The total current $I$ may be obtained by integration over
time of Kirchhoff's equation in the phase representation,
$I=I_{CR}\sin(\phi)+I_{QP}(2eV=\hbar\dot{\phi})$. In this way
the average time-derivative of the phase is found as a function
of the total current through the junction. The result is shown in
Fig. \ref{fig4} for low temperatures $k_{B}T\ll\Delta_{0}$ and
$\phi_{L}=\phi_{R}=0$, i.e. the geometry in which node to node
tunneling appears and $I_{QP}\sim V^{2}$. The I-V
characteristics deviate clearly from the well known RSJ behavior.
For higher temperatures $T$ of the order of the gap $\Delta_{0}$
we have $I_{QP}\sim V$ (see Fig. \ref{fig2}) and the RSJ-behavior
for the I-V curve is recovered.

We now continue with the discussion of the thermodynamic properties
that are described by an effective action for the phase difference
across the junction. This is especially relevant for small mesoscopic
tunnel junctions with a low capacitance $C$ \cite{physrep}.
 From a second order perturbation
expansion in the tunnel matrix elements one finds the following effective
action for the phase difference $\varphi$ across the junction \cite{aes}
\begin{equation}
	S[\varphi]=\int^{\beta}_{0}d\tau \frac{C}{8e^{2}}
	\left(\frac{\partial\varphi(\tau)}{\partial\tau}\right)^{2}
	+\int^{\beta}_{0}d\tau d\tau'
	\cos(\frac{\varphi(\tau)-\varphi(\tau')}{2}) \alpha(\tau-\tau')
	+\int^{\beta}_{0}d\tau d\tau'
	\cos(\frac{\varphi(\tau)+\varphi(\tau')}{2}) \beta(\tau-\tau') \;.
\end{equation}
The kernels $\alpha$ and $\beta$ describe quasi-particle tunneling and
the Josephson coupling respectively. They are given by
\begin{equation}
	\left. \begin{array}{lcl}\alpha(\tau) \\ \\ \beta(\tau) \end{array}
	\right\}=\int\frac{d^{3}k}{(2\pi)^3}\int\frac{d^{3}k'}{(2\pi)^3}
	\left\{ \begin{array}{lcl} G(k,\tau)G(k',-\tau) \\ \\
	\bar{F}(k,\tau)F(k',-\tau) \end{array} \right\}
	\mid T_{k,k'}\mid^{2} \; ,
\end{equation}
where the normal and anomalous Greens functions $G$ and $F$ were introduced.
The usual approximation, in which the Fourier transform of the
tunnel matrix element $T_{k,k'}$ is taken to be independent of
momenta, leads for s-wave superconductors to the standard
expressions in the literature, i.e. both $\alpha$ and $\beta$ decay
exponentially on the time-scale $\Delta^{-1}$. The same approximation
for a d-wave junction will lead to unphysical results.
For instance the beta-term is found to be zero, as the average over
orientations of $\Delta_{k}$ is zero.

To remedy this shortcoming, one again has to retain the direction
dependence of the tunneling matrix elements (\ref{tunn}). The gap
nodes have a pronounced effect on the long time behavior of $\alpha$
and $\beta$. If $\phi_{L}=\phi_{R}=0$, the gap nodes in the two
superconductors are in the same direction (the ``node to node'' geometry)
and the asymptotic behavior of $\alpha(\tau)$ is $\tau^{-3}$.
This corresponds to a low frequency behavior $\sim\omega^{2}\ln\omega$,
which is 'super-ohmic'. An investigation of the $\omega^{2}\ln\omega$
dissipation remains a subject for further study \cite{boz}, and may
be relevant for the phase diagram and superconductor-insulator transition
in thin granular high-T$_C$ films \cite{films}.
For relative orientations without node-to-node tunneling we find exponential
decay, as was also found for s-wave tunnel junctions.

In conclusion we have calculated the full I-V characteristic for a tunnel
junction between two unconventional superconductors by making a new
Ansatz for the direction dependence of the tunnel matrix elements. The
temperature dependence and a strong dependence on relative orientation
is found. For specific orientations the I-V characteristics differ from
the usual RSJ-like behavior. All of our predictions have experimental
consequences and should be verifiable using thin-film junctions on
bicrystal substrates.

\acknowledgements
We would like to thank P. Hadley, G. Sch\"on, and A.~D. Zaikin
for discussions.
We gratefully acknowledge the support by the Swiss Nationalfonds (AvO),
'Sonderforschungsbereich 195' of the DFG (CB and AvO), and
NSF-grant 92-06023 (GTZ).

%
%\noindent
%Figure captions:

\end{document}